\documentclass[12pt,preprint]{aastex}

\newcommand{\beq}{\begin{equation}}
\newcommand{\eeq}{\end{equation}}
\newcommand{\Mdot}{\dot{M}~}

\newcommand{\kms}{\mbox{ km s$^{-1}$}~}

\newcommand{\Mo}{\mbox{M$_{\odot}$}~}
\newcommand{\Moy}{\mbox{M$_{\odot}$ yr$^{-1}$}~}

\newcommand{\Oiii}{O~[{\sc iii}]~}
\newcommand{\Heii}{He~{\sc ii}~}

\shorttitle{PN Dynamics After the Fast Wind} 
\shortauthors{Garc\'{\i}a-Segura et al.}

\begin{document}

\title{The Dynamical Evolution of Planetary Nebulae  \\
        After the Fast Wind}

\author{G. Garc\'{\i}a-Segura\altaffilmark{1}, J. A. L\'opez\altaffilmark{1}, 
W. Steffen\altaffilmark{1}, J., Meaburn\altaffilmark{1,2}, and A. 
Manchado\altaffilmark{3}}

\altaffiltext{1}{Instituto de Astronom\'{\i}a, Universidad Nacional 
Aut\'onoma de M\'exico, Apdo Postal 877, Ensenada 22800, Baja California, 
Mexico; ggs@astrosen.unam.mx; jal@astrosen.unam.mx; wsteffen@astrosen.unam.mx}
\altaffiltext{2}{Jodrell Bank Observatory, University of Manchester,
Macclesfield SK11 9DL, UK; jm@ast.man.ac.uk}
\altaffiltext{3}{Instituto de Astrof\'{\i}sica de Canarias, C/V\'{\i}a 
L\'actea, E-38200 La Laguna, Tenerife, Spain; amt@iac.es}

\begin{abstract}

In this paper we explore the dynamics of ionization bounded planetary nebulae 
after the termination of the fast stellar wind. 
When the stellar wind becomes negligible, the hot, shocked bubble 
depressurizes and the thermal pressure of the photoionized region, 
at the inner edge of the
swept-up shell, becomes dominant. At this stage the shell tends to fragment
creating clumps with  comet-like tails and long, 
photoionized trails in between, while the photoionized material expands
back towards the central stars as a rarefaction wave.
Once that the photoionized gas fills the inner cavity, it develops a kinematical pattern
of increasing velocity from the center outwards with a typical range of velocities starting
from the systemic velocity to $\sim 50 \kms $ at the edges. The Helix nebula is a clear 
example of a planetary nebula at this late evolutionary stage.

\end{abstract}


\section{Introduction}

High spatial resolution imagery in recent years have revealed a number of well developed 
planetary nebulae that exhibit a fragmented appearance of their shells with clumps and radial 
spokes, as it is the case of the famous Helix Nebula, NGC 7293 
(Vorontsov-Velyaminov 1968; 
Meaburn et al. 1992; O'Dell \& Handron 1996). Similar examples are the Dumbbell nebula, NGC 6853 (Manchado et al.1996; Meaburn et al. 2005a) and NGC 2867 (Corradi et al. 2003). Rayleigh-Taylor instabilities have been some times invoked as the cause of this fragmentation but with limited success (Burkert \& O'Dell 1996). Alternatively, the effects of the ionizing radiation on the nebula have been shown by Capriotti (1973) and also by Garc\'{\i}a-Segura \& Franco (1996) and 
Garc\'{\i}a-Segura et al.(1999) to be much more relevant in the fragmentation process of  the nebular shell. 

Recently, Meaburn et al. (2005b) have shown that the current kinematics of nebulae such as NGC 7293 cannot be explained by the classic ``two-wind model'' (Kwok 1982). Since NGC 7293 shows 
emission in \Heii and \Oiii at very low velocities close to the central star, Meaburn et al. (2005b) 
concluded that only after the fast wind has switched off could this global velocity structure be generated.
The absence of a fast stellar wind in NGC 7293 has been confirmed from 
observations with the IUE satellite (Perinotto 1983; Cerruti-Sola \& 
Perinotto 1985) that failed to detect P-Cygni profiles from their central 
stars; likewise in the cases of NGC 6853 and NGC 2867. Furthermore, 
the lack of extended, soft x-ray emission from NGC 7293 (Guerrero et al. 
2001) discards the presence of a "hot bubble" and the  shocked, fast wind. 
In general,  Cerruti-Sola \& Perinotto (1985) found that the hotter the 
central stars are the least likely to have fast stellar winds, a result 
indicative of the rapid fading of the stellar wind as the central star evolves.

In this letter we present  numerical calculations which show that the decay 
of the fast stellar winds produces self-consistently the fragmented structures
and kinematics observed in evolved planetary nebulae.

\section{Relevant Stellar Parameters}

There are several dynamical studies of the evolution of PNs that include
the effects of stellar evolution. Recently,  Sch\"onberner et al. (2005)
have discussed the stellar evolution path for a $0.595$ \Mo post-AGB,
following the recommendation of Pauldrach et al.(1988) for
the parameters of the central stellar wind. A Reimers' wind (Reimers 1975)
is assumed during the short transition region from the AGB to the PN domain.
In this model (see Figure 1 in Sch\"onberner et al.2005), the mass loss rate
begins with several $10^{-7}$ \Moy for a period of $10^3$ yr, decreasing
down to $10^{-8}$ \Moy, to finally  reach values of $10^{-10}$ \Moy
after $10^4$ yr; meanwhile the wind velocity increases from $10^2$ to
$10^4$ \kms. After $10^4$ yr the wind can be considered negligible
(turn-around point).
As initial condition, this calculation assume $4.3 \times 10^{-5}$ \Moy
and $v_{\infty}= 10$ \kms for the AGB wind.

Also, Villaver et al. (2002a) have used the paths from
Vassiliadis \& Wood (1994) for
$0.569$ \Mo, $0.597$ \Mo, $0.633$ \Mo, $0.677$ \Mo, $0.754$ \Mo and
$0.9$ \Mo, corresponding to initial values at the ZAMS of 1, 1.5, 2, 2.5,
3.5 and 5 \Mo respectively. The Vassiliadis \& Wood's (1994) models
also use the Pauldrach et al.(1988) theoretical results to compute the
mass loss rates and wind velocities from the AGB to the PN domain.
In these models (Figure 2 in Villaver et al. 2002a), the fast wind is
negligible (turn-around point) after $30 \times 10^3$, $ 10 \times 10^3$,
$ 5 \times 10^3$, $ 2.5 \times 10^3$, $ 1 \times 10^3$, and
$ 0.2 \times 10^3$ years respectively. The second model in Villaver et
al. (2002a) is almost identical to the model used in
Sch\"onberner et al. (2005).
The initial conditions (see Villaver et al. 2002b) corresponding to the end
of the AGB winds are taken from Vassiliadis (1992),
and are in the range of $4 \times 10^{-6} - 30 \times 10^{-6} $ \Moy,
and $11 - 15$ \kms.

Mellema (1994) has also used a model for a $0.598$ \Mo post-AGB from
Sch\"onberner (1983), with similar stellar parameters as discussed above.

Given the range of possible parameters to consider in the stellar
evolution tracks, stellar masses, mass loss rates and wind velocities,
we have adopted in this study the following representative numbers:
For the AGB winds, we have adopted a $v_{\infty}= 10 \kms$,
$\Mdot = 10^{-6} \Moy$ (model A) and $\Mdot = 10^{-5} \Moy$ (model B).
For the post-AGB wind, $v_{\infty}= 1,000 \kms$ and $\Mdot = 10^{-7} \Moy$.
These choices give wind momentum ratios of 10 (model A) and 1 (model B).
We also have adopted  1,000 yr  for the duration of the fast wind.

\section{Simplified Hydrodynamical Simulations}

The numerical experiment presented here 
(model B)(Garc\'{\i}a-Segura et al.2006 discusses model A)
includes two phases:
in the first phase, a typical two-wind model scenario (Kwok 1982) 
where a fast wind sweeps up a slow wind is considered.
For simplicity, this phase lasts 1,000 yr 
in the computation, but it could last longer or shorter depending of the 
particular track of stellar evolution. In the second phase, the fast wind is 
switched off, and the dynamical evolution is computed for a total of 8,000 yr.
In both phases the photoionization is considered following the approach of 
Garc\'{\i}a-Segura \& Franco 1996 with  a central star that emits 
$10^{45} s^{-1}$ ionizing photons. 
We have adopted this average value from
Villaver et al. (2002a) (see their  Figure 4), which is maintained for at least
$10^4$ yr.
We have not considered in detail the temporal variation of the ionizing 
photons.  However, Villaver's  models show that the evolution of the nebulae 
are optically thick during the first few thousand years of the evolution.
In this regime, our simple models provide a reasonable description of
the location of the ionization front for the mass-loss rates that we have
considered.
A simple expanding spherical morphology  is adopted for simplicity (i. e., no
rotation,  magnetic field, or anisotropic mass-loss events).

The simulations are performed with the hydrodynamical code ZEUS-3D 
(version 3.4) (Stone \& Norman 1992; Clarke 1996), and details about the 
set up can be found in Garc\'{\i}a-Segura et al.(1999), and
Garc\'{\i}a-Segura et al.(2005) for the self-expanding grid technique.

We perform the two-dimensional simulation in spherical polar 
coordinates ($r,\theta$), with reflecting boundary conditions at the 
equator and the polar axis, and rotational symmetry assumed with respect 
to the latter.  
Our grid consist of $200 \times 180$
equidistant zones in 
$r$ and $\theta$ respectively, with a radial extent of 0.1 pc (initially),
and an angular extent of $90^{\circ}$.  The innermost
radial zone lies at $r=2.5 \times 10^{-3}\,$pc from the central star.

Figures 1 and 2 show  mosaics of snapshots of gas densities and photoionized
gas densities covering 8,000 yr of the evolution for
model B, 
each one separated
by 1,000 yr. 
The first, top-left panel in Figure 1 shows progressively, 
starting from the origin of the grid, the fast, free expanding wind region,  
the terminal or reverse shock, the hot, shocked bubble,
the swept-up shell with an outer, forward shock, and the slow wind. This
is the classical view of a two-wind model. Vishniac (1983) or thin
shell instabilities give the corrugate appearance of the swept-up shell
(see Garc\'{\i}a-Segura \& Mac Low 1995 for further details).
The rest of the panels show how the hot shocked bubble collapses after 
$\sim$ 6,000 yr and also show the expansion of the photoionized gas, once that
the fast wind is switched off.
The fragmentation of the swept-up shell follows the 
pattern imposed by the Vishniac instability, and its development and
growth comes from the ionization-shock front instability (Garc\'{\i}a-Segura 
\& Franco 1996).

Figure 3 is a mosaic with snapshots of the radial velocities counterparts of 
Figure 1 and 2.  The last two panels (bottom-right) show the global kinematics
of the ionized gas after the collapse of the hot bubble in which it can be 
clearly appreciated that the gas close to the central star presents 
very small velocities, reaching practically inert conditions there, 
as it is found in the Helix nebula (Meaburn et al.2005b). 
Notice that these low velocities reach the central star region, and are
not only  a projection effect along  the line of sight, as in models where
the low velocity emitting regions are located in the inner edges of the 
swept-up shells (e. g., Marten \&  Sch\"onberner 1991; Perinotto et al. 2004).
The gas
located at the border of the main nebula, shows larger velocities 
following the kinematics of a rarefraction wave. 

The last panel of Figure 3 also shows that the 
neutral cometary globules immersed in the nebula
present smaller velocities than the surrounding photoionized gas 
that has been accelerated by thermal pressure,
while the extended sections of these
neutral tails at further distances show larger velocities than the AGB wind 
due to the acceleration by the ionization-shock front.
Model B is not shown in this letter, but it is discussed in
Garc\'{\i}a-Segura et al. (2006) where similar results are found.

In the simulations, the cometary globules originate in the neutral, swept-up
shell of piled up AGB wind, and so, they, or part of them must be
excited by shocks.  This is consistent with the detection of shock-excited
molecular Hydrogen in the Dumbbell nebula (see Manchado et al. 2006).
Once that the wind is switched off,  the head of the
globules are no longer confined in the swept-up shell by the pressure of the
wind, and so, they become engulfed in the nebula. This is a novel
result that previous hydrodynamical works had not solved (see for example
Garc\'{\i}a-Segura et al. 1999).
Huggins et al.(1992) observed that the CO cometary globules in
the Helix nebula show a
smaller expansion velocity than the ionized gas, which is in  line
with our computations. However, the present numerical simulations are two-dimensional,
and the limited resolution does not allow a quantitative analysis of the structures that are formed.
Thus, quantitative details about velocities of the globules, number, 
sizes and masses, as well as 
the spoke-like pattern and its spatial 
frequency developed in the simulations are beyond the scope of this letter. 
However, the qualitative resemblance with cases like the Helix and
the Dumbbell nebulae is remarkable and encouraging. Note for example, the
cometary globules that develop in Figure 2, bottom-right, with bright 
photoionized heads,
like the ones observed in the Helix
(see also Figure 16 in O'Dell et al.  2004), and also the resemblance with
the molecular, shock-exited long tails extending further out in the
Dumbbell (Manchado et al.2006).

In this study, we have switched off the fast wind after the first 1000 years of evolution and followed the subsequent development finding for the first 
time a consistent 
representation of the fractured morphology and kinematic pattern often observed in planetary 
nebulae. The stellar wind is likely not to die-off as abruptly as in our 
simulations, although
massive central stars evolve fast (e.g.  see fig 3 in Villaver et al. 2002) 
and the behavior computed here should be a 
good approximation 
for those cases, as in the Helix nebula where the estimated central star 
mass is 0.93 \Mo (G\'orny et al. 1997).
For the case of the evolution of low mass central stars, those are 
expected to take 
place in much longer times  than the dilution of their surrounding nebulae. 
If this is the case, there should be a direct  correlation between the
fragmentation of a planetary nebula with the mass of the central star, 
and an anticorrelation with the detection of extended soft x-rays. 

Additional numerical simulations at higher resolution, taking detailed 
account of the time dependent stellar evolution effects will allow a 
quantitative analysis of the shell structure as a result of the 
termination of the fast wind. Such calculations will be the subject of  
a future paper.

\acknowledgments
We thank the anonymous referee for his/her valuable comments which 
improved the manuscript.
We thank M. L. Norman and the LCA for the use of ZEUS-3D. 
The computations were performed at 
Instituto de Astronom\'{\i}a, Universidad Nacional Aut\'onoma de M\'exico. 
We gratefully acknowledge financial support from UNAM-DGAPA grants  IN108406-2, IN108506-2 and IN112103 as well as CONACYT grant 43121

\clearpage

\begin{figure}
\epsscale{.80}
\caption{Snapshots of gas densities covering 8,000 yr of the evolution of 
model B}
\end{figure}

\clearpage

\begin{figure}
\caption{Same as Figure 1 but for photoionized gas densities. The white
line at $45^{\circ}$ in the last panel correspond to the one-dimensional plot
on figure 3}
\end{figure}

\clearpage

\begin{figure}
\caption{Same as Figure 1 but for radial velocities. A one-dimensional plot
at $45^{\circ}$ is also shown in the last panel at 8,000 yr.}
\end{figure}

\end{document}